\documentstyle[12pt]{article}
\begin{document}
\begin{titlepage}
\begin{center}
{\LARGE {
BPS Solutions in D=5 \\
\vskip 0.3cm
Dilaton-Axion Gravity
}}\\
\vskip 1.5cm
{  {\bf Oleg Kechkin$^1$
and
Maria Yurova$^2$ }} \\
\vskip 0.5cm
{\small
Institute of Nuclear Physics,
Moscow State University, \\
Vorob'jovy Gory, Moscow 119899, RUSSIA. \\
}
\end{center}
\vfill
\begin{center} {\bf Abstract}
\end{center}
{
\small
We show that the D=5 dilaton-axion gravity compactified on a 2-torus 
possesses the SL(4,R)/SO(4) matrix formulation. It is used for
construction of the SO(2,2)-invariant BPS solution depended on the one
harmonic function.
}
\vspace{2mm} \vfill \hrule width 3.cm
{\footnotesize
\noindent
$^1$ kechkin@monet.npi.msu.su \\
$^2$ yurova@depni.npi.msu.su}
\end{titlepage}
\newpage
\section {Introduction}
Low energy limit of superstring theories provides some modifications of
general relativity possessing non-trivial symmetry groups and different
matrix formulations. So, the bosonic sector of heterotic string theory,
which describes by the $D=d+3$ dilaton-axion gravity coupled to the $n$
Abelian vector fields, allows the SO(d+1,d+n+1) symmetry after the
compactification on a $d$-torus \cite {ms}-\cite {s}. Its
SO(d+1,d+n+1)/S[O(d+1)$\times$ O(d+n+1)]
chiral formulation was extensively used for construction of exact
solutions and for general analysis of the low energy heterotic string states,
especially in the critical case of d=7 and n=16.

In this letter we consider the d=2, n=0 theory, which also arises in the
frames of the D=5 supergravity \cite {gst}, and can be considered as
its special electromagnetic-free case \cite {sv}-\cite {gkltt}. We show that
this theory
allows the SL(4,R)/SO(4) formulation in addition to the known one
SO(3,3)/S[O(3)$\times$O(3)]. The similar situation takes place for the
model with with d=n=1, when both the SO(2,3)/S[O(2)$\times$O(3)] and
Sp(4,R)/U(2) representations become possible.

Next, following the Kramer-Neugebauer approach \cite {kn}, developed
in \cite {cl} for the D=5 Kaluza-Klein theory and in \cite {ky}-\cite {gcl}
for
the d=n=1 system, we study the backgrounds trivial at space infinity and
depended on the one harmonic function. It is shown that the corresponding
solutions of motion equations can be obtained from the trivial one (i.e. from
the solution describing the empty and flat D=5 space-time) using some
matrix operator of the coset SL(4,R)/SO(2,2). This operator is parametrized
by the full set of physical charges; the group SO(2,2), being the maximal
isometry subgroup preserving a vacuum state, form the general symmetry
group of the solutions under consideration.

The constructed matrix operator defines the general BPS solution in a
parametric form. We present explicit formulae in the case of Coulomb
dependence of this operator on the space coordinates. The corresponding
solution describes the BPS interacting point sources and contains the
black hole case as well the naked singularity one.

Some classes of D=5 BPS solutions with non-trivial values of electromagnetic
charges were obtained in \cite {cfgk}.
\section{\bf New Chiral Formulation} 
In this letter we study the D=5 dilaton-axion gravity described by the
action
\begin{eqnarray}
S^{(5)} = \int \mid G^{(5)} \mid ^{\frac {1}{2}} e^{-\phi ^{(5)}}
\left \{ R^{(5)} + \phi ^{(5)}_{;M} \phi ^{(5);M} - 
\frac {1}{12}H^{(5)}_{MNK}H^{(5)MNK}\right \},
\end{eqnarray}
where $\phi ^{(5)}$ is dilaton and the axion $H^{(5)}_{MNK}$ is related
with the antisymmetric Kalb-Ramond field $B_{NK}$ (M, N, K = 1, ..., 5) as
follows:
\begin{eqnarray}
H^{(5)}_{MNK} = \partial _{M} B_{NK} + \partial _{K} B_{MN} +
\partial _{N} B_{KM}.
\nonumber
\end{eqnarray}

The compactification of this theory on a 2-torus was performed in
\cite {ms}-\cite {s}.
The complete set of the 3-dimensional variables includes 
\vskip 0.7cm
\noindent
two $2 \times 2$ matrices $G = [G_{pq}]$ and $B = [B_{pq}]$ \, (p, q = 1,2), 
\begin{eqnarray}
G_{pq} = G^{(5)}_{p+3, q+3}, 
\qquad
B_{pq} = B^{(5)}_{p+3, q+3};
\end{eqnarray}

\noindent
two $2 \times 3$ matrices $U = [U_{p\mu}]$ and $V = [V_{p\mu}]$ \, 
$(\mu, \nu = 1,2,3)$,
\begin{eqnarray}
U_{p\mu} = (G^{-1})_{pq}G^{(5)}_{p+3, \mu},
\qquad
V_{p\mu} = B^{(5)}_{p+3, \mu} - B_{pq}U_{q\mu};
\end{eqnarray}

\noindent
and also the 3-dimensional dilaton, metric and Kalb-Ramond fields 
\begin{eqnarray}
\phi = \phi ^{(5)} - \frac {1}{2} \, {\rm ln \, det} G,
\end{eqnarray}       
\begin{eqnarray}
g_{\mu \nu} = e^{-2\phi} \left [ G^{(5)} - 
U^T G U \right ]_{\mu \nu},
\end{eqnarray}
\begin{eqnarray}
B_{\mu \nu} = \left [ B^{(5)} - 
U^T B U  - \frac {1}{2} \left ( U^T V - V^T U \right ) 
\right ]_{\mu \nu}
\end{eqnarray}       
(The quantity $B_{\mu \nu}$ can be taken equal to zero in view of its
non-dynamic properties in three dimensions).

This system allows the dualization on shell. Namely, one can introduce the
3-dimensional pseudoscalar columns $u = [u_p]$ and $v = [v_p]$ accordingly
to
\begin{eqnarray}
\nabla u = e^{-2\phi}\left [ G\nabla \times \vec U -
BG^{-1}(\nabla \times \vec V + B \nabla \times \vec U) \right ],
\end{eqnarray}       
\begin{eqnarray}
\nabla v = e^{-2\phi}G^{-1} \left [ \nabla \times \vec V + 
B \nabla \times \vec U \right ].
\end{eqnarray}       
Here the $2 \times 1$ vector columns $\vec U$ and $\vec V$ are constructed
of the components $\vec U_p = (U_{p \mu})$ and $\vec U_p = (U_{p \mu})$
correspondingly and all 3-dimensional vector operations are related with
the metric $g_{\mu \nu}$.

The resulting model becomes chiral one with the action
\begin{eqnarray}
S = \int d^3 x g^{\frac {1}{2}} \{ 
- R + \frac {1}{8}{\rm Tr}(J^{{\cal M}})^2
\},
\end{eqnarray}
where $J^{{\cal M}} = \nabla {\cal M} \, {\cal M}^{-1}$. The
symmetric matrix ${\cal M}$
\begin{eqnarray}
{\cal M} = \left (\begin{array}{crc}
{\cal G}^{-1}&\quad&{\cal G}^{-1}{\cal B}\\
-{\cal B}{\cal G}^{-1}&\quad& {\cal G} - {\cal B}{\cal G}^{-1}{\cal B}\\
\end{array}\right )
\end{eqnarray}
has the following $3\times 3$ block components \cite {hk}:
\begin{eqnarray}
{\cal G} = \left (\begin{array}{crc}
- e^{-2\phi} + v^T G v &\quad& v^T G\\
G v&\quad& G\\
\end{array}\right ), \quad
{\cal B} = \left (\begin{array}{crc}
0 &\quad& - (u+Bv)^T\\
u+Bv &\quad& B\\
\end{array}\right ).
\end{eqnarray}
It possesses the $SO(3,3)$ group property
\begin{eqnarray}
{\cal M}^T {\cal L} {\cal M} = {\cal L}, \quad
{\rm where} \quad
{\cal L} = \left (\begin{array}{crc}
0 &\quad& I_3\\
I_3 &\quad& 0\\
\end{array}\right ),
\end{eqnarray}
where $I_3$ is the $3 \times 3$ unit matrix, so {\cal M} $\in$
SO(3,3)/S[O(3)$\times$O(3)]. The other form of this matrix
was presented by Sen in \cite {s} for the case of d=7, n=16.

Two group isomorphisms, SO(3,3)$\sim$SL(4,R) and S[O(3)$\times$O(3)]
$\sim$ SO(4), provide an existence of the new SL(4, R)/SO(4) chiral
formulation of a problem. Actually, let us introduce the $3 \times 1$
column ${\cal H}$ 
\begin{eqnarray}
{\cal B}_{mn} = \epsilon _{mnk} {\cal H}_k,
\end{eqnarray}
where $m, n, k = 1,2,3$ and $\epsilon _{mnk}$ is the antisymmetric 
tensor with $\epsilon _{123} = 1$. Then, after some algebraic
manipulations one obtains from Eq. (9) that
\begin{eqnarray}
S = \int d^3 x g^{\frac {1}{2}} \{ 
- R + \frac {1}{4}{\rm Tr} (J^{{\cal N}})^2
\},
\end{eqnarray}
where $J^{{\cal N}} = \nabla {\cal N} \, {\cal N}^{-1}$ and ${\cal N}$
is given by 
\begin{eqnarray}
{\cal N} = {({\rm det} {\cal G})}^{- \frac {1}{2}}\left (\begin{array}{crc}
{\cal G} &\quad& {\cal G}{\cal H}\\
{\cal H}^T {\cal G} &\quad& {{\rm det} {\cal G}} + 
{\cal H}^T {\cal G}{\cal H}\\
\end{array}\right ).
\end{eqnarray}

It is easy to prove that the symmetric matrix $\cal N$ is unimodular, so
this matrix actually belongs to SL(4,R)/SO(4). This new chiral formulation
is the simplest one in view of its trivial group property and low matrix
dimension. We use it in the next section for the analysis of asymptoticaly
flat heterotic string backgrounds.

The SL(4,R)/SO(4) coset formulation arose previously in the frames of the
D=6 Kaluza-Klein theory compactified
on a 3-torus \cite{cy}. Such isomorphism of two different theories has not
any analogy for other values of $d$ and $n$.
\section {Harmonic Solution}
The equations of motion
\begin{eqnarray}
\nabla \left ( J^{\cal N} \right ) = 0, \qquad
R_{\mu \nu} = \frac {1}{4} {\rm Tr} \, \left ( 
J^{\cal N}_{\mu}J^{\cal N}_{\nu}
\right ),
\end{eqnarray}
allow the solution anzats \cite {} ${\cal N} = {\cal N}(\lambda)$
where the coordinate function $\lambda (x^i)$ satisfies to the Laplace
equation
\begin{equation}
\nabla ^2 \lambda = 0.
\end{equation}
From the matter part of Eq. (16) it follows that
\begin{equation}
{\cal N} = e^{\lambda T}{\cal N}_0 \equiv {\cal S}{\cal N}_0,
\end{equation}
where ${\cal N}_0 = {\cal N}|_{\lambda = 0}$ and $T$ is the constant
matrix. The Einstein part of Eq. (16) transforms to 
\begin{eqnarray}
R_{\mu \nu} = \frac {1}{4} {\rm Tr} \left ( T^2 \right )
\nabla _{\mu} \lambda \nabla _{\nu} \lambda,
\end{eqnarray}
so both the equations (17) and (19) correspond to the action
\begin{eqnarray}
S = \int d^3 x g^{\frac {1}{2}} \left \{ 
- R + \frac {1}{4}{\rm Tr} (T^2) \, \left ( \nabla \lambda \right ) ^2
\right \}.
\end{eqnarray}

We define
\begin{eqnarray}
{\cal N}_0 = \left (\begin{array}{crc}
- I&\quad&0\\
0&\quad&I 
\end{array}\right ),
\end{eqnarray}
where block matrices are $2 \times 2$ ones. The corresponding quantities
$\phi _0, \, u_0, \, v_0, \, B_0$ are trivial and $G_0 = {\rm diag} \,
(-1, 1)$, so this field configuration describes the empty and flat
5-dimensional space-time.

To provide the SL(4,R)/SO(4) coset property of ${\cal N}$, one must restrict
the operator ${\cal S}$ by the relations ${\rm det} \, {\cal S} = 1$ and
${\cal S}^T = {\cal N}_0 {\cal S} {\cal N}_0$. Then for the matrix $T$ one
obtains:
\begin{eqnarray}
{\rm Tr} \, T = 0, \quad {\rm and} \quad T^T = {\cal N}_0
T {\cal N}_0.
\end{eqnarray}
Using these formulae one can establish after some algebraic calculations
that the matrix T satisfies to the relation
\begin{eqnarray}
T^4 = \frac {1}{4}{\rm Tr}(T^4) + \frac {1}{2}{\rm Tr}(T^2)\,[T^2 -
\frac {1}{4}{\rm Tr}(T^2)] + \frac {1}{3}{\rm Tr}(T^3)T.
\end{eqnarray}
This property is crucial for the explicit calculation of the exponential
${\cal S} = e^{\lambda T}$. Namely, let us denote four roots of
the equation
\begin{eqnarray}
k^4 = \frac {1}{4}{\rm Tr}(T^4) + \frac {1}{2}{\rm Tr}(T^2)\,[k^2 -
\frac {1}{4}{\rm Tr}(T^2)] + \frac {1}{3}{\rm Tr}(T^3)k, 
\end{eqnarray}
as $k_{\alpha}$, where $\alpha = 1,2,3,4$. Then for the different
roots the exponentiation result is given by 
\begin{eqnarray}
&&{\cal S} = \sum _{\alpha} e^{k_{\alpha}\lambda}
\{k_{\alpha}^3 - \frac {1}{2}{\rm Tr}(T^2)k_{\alpha} -
\frac {1}{3}{\rm Tr}(T^3) +
[k_{\alpha}^2 - \frac {1}{2}{\rm Tr}(T^2)]T +
\nonumber
\\
&&k_{\alpha}T^2 + T^3\} \prod _{\beta \neq \alpha} (k_{\alpha} -
k_{\beta})^{-1},
\end{eqnarray}
while the formulae for various cases of coinciding roots can be obtained
from Eq. (25) using corresponding limes procedure. The exceptional case
of zero roots is related with zero traces, ${\rm Tr}(T^2) = {\rm Tr}(T^3) =
{\rm Tr}(T^4) = 0$. This leads to the vanishing value of $T^4$ (see Eq.(23)),
so for the operator under consideration one has ${\cal S} = 1 + \lambda T +
\lambda ^2T^2/2 + \lambda ^3T^3/6$.

Using Eq. (24) it is easy to express three independent traces in terms
of the roots $k_{\alpha}$:
\begin{eqnarray}
&&{\rm Tr}(T^4) - \frac {1}{2}[{\rm Tr}(T^2)]^2 =
4\prod _{\alpha}k_{\alpha}, \quad
{\rm Tr}(T^3) = 3\prod _{\alpha} k_{\alpha} \sum _{\beta}
k_{\beta}^{-1}, \nonumber \\
&&{\rm Tr}(T^2) = - \sum _{\beta \neq \alpha}k_{\alpha}k_{\beta}.
\end{eqnarray}
The reversed relations (roots as functions of traces) are more
complicated. These ones can be obtained using the Cardano and Ferrari 
formulae; we will not write them here.

Thus, original theory reduces to the system (20) in the frames
of the anzats under consideration. This means that any solution of this
system, which coincides with the 3-dimensional Einstein-Klein-Gordon model,
can be transformed to the solution of the theory (1) using the operator 
${\cal S}$ defined by Eq. (25).

Now let us establish symmetries of the discussing backgrounds. It is easy
to see that the matrix $\Gamma$, defined by the relation
${\Gamma}^T = - {\cal N}_0{\Gamma}{\cal N}_0$, is the generator of symmetry
transformation preserving the vacuum value ${\cal N}_0$. Actually, the
matrix ${\cal C} = e^{\Gamma}$ belongs to $SL(4,R)$ and can be considered as
the operator of symmetry transformation ${\cal N} \rightarrow {\cal C}^T
{\cal N}{\cal C}$. Next, it satisfies to the relation ${\cal C}^T
{\cal N}_0{\cal C} = {\cal N}_0$, i.e. moreover ${\cal C}\in SO(2,2)$. For
the anzats under consideration this transformation is equivalent to the map
$T \rightarrow {\cal C}^T T {{\cal C}^T}^{-1}$ which preserves the algebraic
property (12). Thus, it defines the SO(2,2) reparametrization of the
matrix T. Finally, one can see that T $\in$ sl(4,R)/so(2,2), thus
{\cal S} $\in$ SL(4,R)/SO(2,2).
\section{BPS States}
The effective 3-dimensional system (20) is equivalent to the static D=4
Einstein theory if ${\rm Tr}(T^2)>0$; the Einstein's metric element
$g_{tt}$ is related with our harmonic function $\lambda$ as $g_{tt} =
\sqrt {{\rm Tr}(T^2)/2}{\rm ln}|\lambda|$.
Below we consider backgrounds saturated to the Bogomol'nyi-Gibbons-Hull
bound ${\rm Tr}(T^2)=0$ which are related with BPS states of the original
heterotic string theory. In this case the 3-metric $g_{\mu \nu}$ becomes
flat, so
one can put $ds^2_{(3)} = d\vec r ^2$ without loss of generality. Next,
the trivial at space infinity solution of Eq. (17) reads $\lambda =
\sum_k \lambda _k/|\vec r - \vec r_k|$, where $\lambda_k = const$. The
resulting field
configuration describes the system of BPS sources located at the points
$\vec r_k$ (see \cite {} for the details in the case of d=n=1 theory).

Let us define the background charges as ($r \rightarrow \infty$)
\begin{eqnarray}
\phi \rightarrow \frac {{\cal D}}{r}, \quad B \rightarrow 
\frac {2K}{r}\sigma_2, \quad
u \rightarrow \frac {2N_u}{r}, \quad v \rightarrow \frac {2N_v}{r}, \quad
G \rightarrow - \left ( 1 - \frac {2M}{r} \right ) \sigma_3. 
\end{eqnarray}
Then, comparing these formulae with the equation ${\cal N} \rightarrow
(1 + T/r){\cal N}_0$, one can find the following relationship between
block components of $T$ and introduced charges:
\begin{eqnarray}
T =
\left (\begin{array}{ccc}
-{\cal D}+{\rm Tr}M & 2N^T_v & 2K \cr
2\sigma_3N_v & {\cal D}+{\rm Tr}M - 2M & -2\sigma_1N_u \cr
-2K & 2N_u\sigma_2 & -{\cal D}-{\rm Tr}M
\end{array}\right ). 
\end{eqnarray}
Thus, the matrix $T$ is the charge matrix of the theory.

Now let us consider the special solution corresponded
to the case of single BPS source with $T^2 = 0$. Then $\lambda = 1/r$ and
\begin{eqnarray}
ds^2_{(3)} = dr^2 + r^2(d\theta ^2 + \sin ^2 \theta \, d\varphi ^2).
\end{eqnarray}
For the potentials $u$ and $v$ one obtains:
\begin{eqnarray}
u = 2\left ( 1 - \frac {2M}{R}\right ) \frac {RN_u}{\Delta}, \quad
v = 2\left ( 1 + \frac {2\sigma_2M\sigma_2}{R}\right )\frac {RN_v}{\Delta},
\end{eqnarray}
where $R = r+{\cal D} + {\rm Tr} M$ and $\Delta = (R - {\rm Tr}M)^2 -
\left ( {\rm Tr} M \right )^2 + 4{\rm det} M$, while the extra
components of the Kalb-Ramond field read
\begin{eqnarray}
B^{(5)}_{p+3, q+3} = \frac {2\sigma_2}{(R-{\cal D} - {\rm Tr} M)^2}
\left \{
\frac {K\Delta}{R} +
2N^T_v\sigma_1\left (1 - \frac {2M}{R}\right )
N_u \right \}.
\end{eqnarray}
Next, the 3-dimensional dilaton function and extra metric components are
\begin{eqnarray}
e^{2\phi} = \frac {\Delta}{(R-{\cal D} - {\rm Tr} M)^2},
\qquad G = - \left ( 1 - \frac {2M}{R}
\right ) \sigma_3,
\end{eqnarray}
so the 5-dimensional dilaton takes the form
\begin{eqnarray}
e^{\phi^{(5)}} = \frac {\Delta}{R(R-{\cal D} - {\rm Tr} M)}.
\end{eqnarray}
Thus, the physical 5-dimensional dilaton charge is ${\cal D}^{(5)} =
{\cal D} - {\rm Tr} M$. Finally, from Eqs. (2), (3), (5) and (7) it follows
that the 5-dimensional line element reads
\begin{eqnarray}
ds^2_{(5)} = G_{pq}(dx -\sigma_3N_u\cos \theta d\varphi )^{3+p}(dx -
\sigma_3N_u\cos \theta d\varphi )^{3+q} + e^{2\phi}ds^2_{(3)}.
\end{eqnarray}

Charges of the solution (29)-(34) are not independent: the matrix relation
$T^2 = 0$ puts the following five scalar constraints on the set of nine
charge parameters:
\begin{eqnarray}
&&{\cal D}{\rm Tr}(M\sigma_3) + N^T_u\sigma_3N_u - N^T_v\sigma_3N_v =0, \quad
{\cal D}{\rm Tr} M + N^T_uN_u - N^T_vN_v =0, \nonumber \\
&&2N^T_v\sigma_3N_v + 2N^T_u\sigma_3N_u + {\cal D}^2 - 2K^2 +
{\rm Tr}M^2 = 0, \quad
{\rm det}M - K^2 =0, \nonumber \\
&&{\cal D}\left [{\rm Tr}(M\sigma_2) - 2K\right ] +
(N_v + N_u)^T\sigma_1(N_v + N_u)=0.
\end{eqnarray}
These constraints can be solved explicitly; the resulting solution is
four-parametric and contains both the black hole (with horizon located
at $R_H = {\cal D} + {\rm Tr}M$) and massless naked singularity
branches. It is easy to establish that black-hole backgrounds are
two-parametric and correspond to the additional constraint $N_u = 0$.
\section{Conclusion}
Thus, the low-energy D=5 heterotic string theory allows the SL(4,R)/SO(4)
formulation after compactification on a 2-torus. We have used this one
for the general analysis of backgrounds trivial at the space infinity
and depended on the one harmonic function.

It is shown that the corresponding solutions of motion equations are
invariant under the action of the SO(2,2) group of transformations which
is the maximal subgroup preserving the asymptotic flatness property of
backgrounds. The obtained formulae represent the class of BPS solutions in
a parametric form; the explicit formulae of all potentials are given in the
special case of Coulomb dependence of the chiral matrix on the space
coordinates.

Using the technique developed in this letter it is possible to
construct rotating solutions defined by two harmonic functions, as it had
been done for the D=5 Kaluza-Klein \cite {cl} and d=n=1 low energy heterotic
string theories \cite {gcl}.

The D=5 system without electromagnetic fields can be mapped into the
theory with non-trivial Maxwell sector using the ``charging'' matrix
Harrison transformation. To perform it one can use the Ernst matrix
potential technique developed in \cite {hk}.
\section*{Acknowledgments}
We would like to thank our colleagues for an encouraging relation to
our work.

\end{document}